\documentclass[
amsmath,amssymb,
aps,
prl, reprint,
]{revtex4-1}

\usepackage{amssymb,bm}
\usepackage{graphicx}

\usepackage[%
colorlinks=true,
urlcolor=blue,
linkcolor=blue,
citecolor=blue
]{hyperref} 
\usepackage{amsmath}
\usepackage{epstopdf}
\usepackage{float}
\allowdisplaybreaks

\usepackage[usenames]{color}

\begin{document}
	
	\title{\textbf{Spin splitting of surface states in HgTe quantum wells}}
	
	\author{A.A.\,Dobretsova$^{1, 2}$, Z.D.\,Kvon$^{1, 2}$, S.S.\,Krishtopenko$^{3,4}$, N.N.\,Mikhailov$^1$, S.A.\,Dvoretsky$^1$}
	\affiliation{$^1$Rzhanov Institute of Semiconductor Physics, Novosibirsk 630090, Russia}
	\affiliation{$^2$Novosibirsk State University, Novosibirsk 630090, Russia}
	\affiliation{$^3$Institute for Physics of Microstructures RAS, GSP-105, 603950, Nizhni Novgorod, Russia}
	\affiliation{$^4$Laboratoire Charles Coulomb, UMR CNRS 5221, University of Montpellier, 34095 Montpellier, France.}
	
	\date{\today}
	
	\begin{abstract}
		We report on beating appearance in Shubnikov--de Haas oscillations in conduction band of 18--22\,nm HgTe quantum wells under applied top-gate voltage. Analysis of the beatings reveals two electron concentrations at the Fermi level arising due to Rashba-like spin splitting of the first conduction subband $H1$. The difference $\Delta N_s$ in two concentrations as a function of the gate voltage is qualitatively explained by a proposed toy electrostatic model involving the surface states localized at quantum well interfaces. Experimental values of $\Delta N_s$ are also in a good quantitative agreement with self-consistent calculations of Poisson and Schr\"odinger equations with eight-band \textbf{k$\cdot$p} Hamiltonian. Our results clearly demonstrate that the large spin splitting of the first conduction subband is caused by surface nature of $H1$ states  hybridized with the heavy-hole band.
		\begin{description}
			\item[PACS numbers]
			\pacs{1} 73.21.Fg, \pacs{2} 73.63.Hs, \pacs{3} 73.20.At, \pacs{4} 73.61.Ga
			\item[key words]{spin splitting, Rashba effect, surface states, Shubnikov--de Haas oscillations, quantum wells}
		\end{description}
	\end{abstract}	
	
	\maketitle
	
	\section{Introduction}
	
Thin films based on HgTe are known by a number of its unusual properties originating from inverted band structure of HgTe~\cite{Konig2007,Kvon2008SM1,Buttner2011a,Brune2011}. The latter particularly results in existence of topologically protected gapless states, arising at HgTe boundaries with vacuum or materials with conventional band structure. Although these states were theoretically predicted more than 30 years ago~\cite{Dyakonov1981eng,Dyakonov1982,Volkov1985eng}, clear experimental confirmation was not possible at that time due to lack of growth technology of high quality HgTe-based films. Experimental investigations of wide (the width $d\geqslant70$\,nm) strained HgTe quantum wells (QWs), which started only in 2011, confirmed existence of the predicted surface states and revealed their two-dimensional (2D) nature~\cite{Brune2011,Kozlov2014,Kozlov2016}.

In comparison with other materials with the inverted band structure, in which the surface states are known being Dirac-like~\cite{Xia2009,Hsieh2009Nature,Chen2009}, HgTe spectrum involves heavy-hole band $|\Gamma_8,\pm3/2\rangle$  modifying the surface state dispersion. Although strain opens a bulk band-gap and results thus in three dimensional (3D) topological insulator state of wide HgTe quantum wells~\cite{Brune2011,Kozlov2014,Kozlov2016}, it does not cancel strong hybridization of the surface states with the $|\Gamma_8,\pm3/2\rangle$ band. As a result, the surface states in strained HgTe films can be resolved only at large energies, while at the low ones they are indistinguishable from conventional heavy-hole states~\cite{Gerchikov1990,Dobretsova2015}.

In thin films of 3D topological insulator the surface states from the opposite boundaries may be coupled by quantum tunneling, so that small thickness-dependent gap is opened up~\cite{Linder2009,Liu2010,Lu2010}. In strained HgTe thin films, the latter arises deeply inside the heavy-hole band at the energies significantly lower than the top of the valence band~\cite{Brune2011}. In the ultrathin limit, the HgTe quantum well transforms into semimetal~\cite{Kvon2008SM1,Olshanetsky2009} and then to 2D topological insulator~\cite{Bernevig2006,Konig2007}  with both gapped surface and quantized bulk states.

On the other hand, the electronic states in HgTe QWs are classified as hole-like $H_n$, electron-like $E_n$ or light-hole-like $LH_n$ levels according to the dominant contribution from the bulk $|\Gamma_8,\pm3/2\rangle$, $|\Gamma_6,\pm1/2\rangle$ or $|\Gamma_8,\pm1/2\rangle$ bands at zero quasimomentum $k=0$~\cite{Bernevig2006}. The strong hybridization in inverted HgTe QWs results in the upper branch of the gapped surface states being represented by the $H_1$ subband~\cite{Brune2011}. At large quasimomentum $k$ the wave-functions of $H_1$ subband are localized at the QW interfaces, while at $\Gamma$ point of the Brillouin zone they are localized in the QW center and are thus indistinguishable from other 2D states.

The gapped surface states in the films of 3D topological insulators exhibit sizable Rashba-type spin splitting, arising due to electrical potential difference between the two surfaces~\cite{Shan2010}. Such spin splitting was first observed in QWs of Bi$_2$Se$_3$~\cite{Zhang2009}, which is a conventional 3D topological insulator with Dirac-like surface states~\cite{Zhang2009,Xia2009,Hsieh2009Nature,Chen2009}. The spin splitting of the gapped surface states also exists in HgTe QWs and should be naturally connected with the splitting of the $H_1$ subband. Previous experimental studies of 12--21\,nm wide HgTe QWs~\cite{Zhang2001,Gui2004,Spirin2010} have attributed large spin splitting of the $H_1$ subband to the Rashba mechanism in 2D systems~\cite{Rashba1960,Zawadzki2004}, enhanced by narrow gap, large spin-orbit gap between the $|\Gamma_8,\pm1/2\rangle$ and $|\Gamma_7,\pm1/2\rangle$ bands, and the heavy-hole character of the $H_1$ subband. The latter however contradicts the fact that the splitting of other subbands $H_2$, $H_3$, $H_4$ \textit{etc.} with the heavy-hole character is significantly lower.

In this work, we investigate spin splitting of conduction band in 18--22\,nm HgTe QWs with asymmetrical potential profile tuned by applied top gate voltage. The beating pattern of Shubnikov--de Haas (ShdH) oscillations, observed in all the samples at the applied top gate voltage, reveals two electron concentrations at the Fermi level due to the spin splitting of the $H_1$ subband. Experimental difference in the concentrations as a function of the gate voltage is qualitatively explained by a proposed toy electrostatic model involving the surface states at the QW interfaces. Self-consistent Hartree calculations based on eight-band \textbf{k$\cdot$p} Hamiltonian~\cite{Krishtopenko2016}, being in good quantitative agreement with the experimental data, clearly show that the large Rashba-like spin splitting of the $H1$ subband is caused by the surface nature of $H1$ states hybridized with the heavy-hole states.
	
	\begin{figure}[b]
		\begin{minipage}[h]{1\linewidth}
			\begin{tabular}{p{0.3\linewidth}p{0.8\linewidth}}
				(a) & (b) \\
			\end{tabular}
		\end{minipage}
		\begin{minipage}[h]{0.3\linewidth}
			\centering\includegraphics[scale=1]{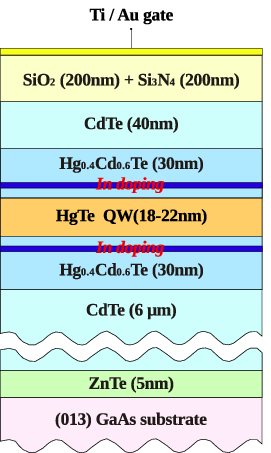} 
		\end{minipage}
		\begin{minipage}[h]{0.68\linewidth}
			\centering\includegraphics[scale=1]{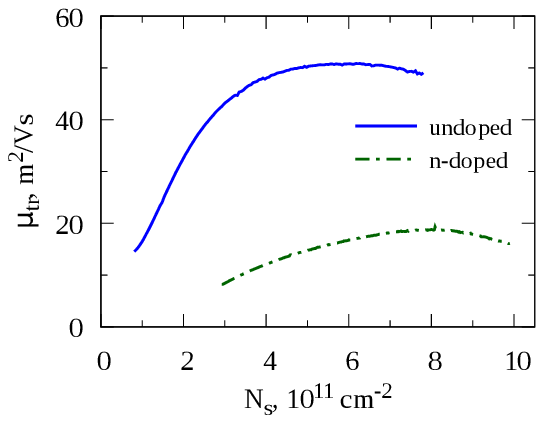} 
		\end{minipage}
		\caption{(a)~The cross section of the structures studied. (b)~Transport mobility dependence on electron concentration for undoped (\#081112) and symmetrically \textit{n}-doped (\#130213) samples.}
		\label{pic:sample}
	\end{figure}
	
	\section{Experiment}
	
	\begin{figure}[t]
		\flushleft\includegraphics[scale=1.2]{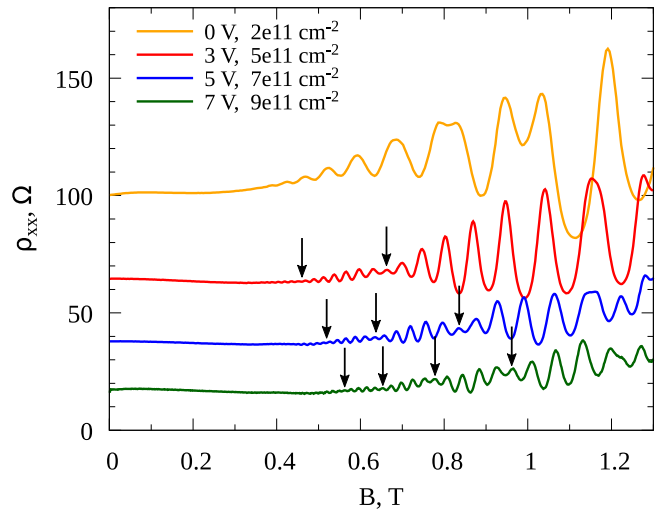} \\
		\caption{Longitudinal resistivity $\rho_{xx}$ dependences on magnetic field $B$ at top gate voltages $V_g=0-7$\,V obtained for undoped 22\,nm HgTe quantum well \#081112. Arrows indicate oscillation beatings.}
		\label{pic:081112_ROxx}
	\end{figure}
	
	Our experiments were carried out on undoped 22\,nm (\#081112) and symmetrically \textit{n}-doped 18\,nm (\#130213) HgTe quantum wells with (013) surface orientation. The samples were grown by molecular beam epitaxy, the detailed description of their preparation can be found in~\cite{Mikhailov2006a,Kvon2011rev}. The cross section of the structures is shown in Fig.\,\ref{pic:sample}\,(a). The structures were patterned into Hall bars with metallic top gate, distances between the contacts 100 and 250\,$\mu$m and the bar width 50\,$\mu$m. Electron concentration of \textit{n}-doped sample \#130213 at zero gate voltage was $N_s = 7.3\times 10^{11}$\,cm$^{-2}$. The experiments were performed at temperatures from 2 to 0.2\,K and magnetic fields up to 8\,T. For magnetotransport measurements the standard lock-in technique was used with the excitation current 100\,nA and frequencies 6\,-\,12\,Hz. In this study we were interested in electron transport when only the first conduction subband is occupied. Electron concentration was thus in the range $1-9\times10^{11}$\,cm$^{-2}$. The electron mobility in this region was rather high (see Fig.\,\ref{pic:sample}\,(b)) within $10-60$\,m$^2$\,/V\,s for undoped and $8-20$\,m$^2$\,/V\,s for doped samples.
	
	\begin{figure}
		\begin{minipage}[h]{1\linewidth}
			\begin{tabular}{p{0.4\linewidth}p{0.7\linewidth}}
				(a) & (b) \\
			\end{tabular}
		\end{minipage}
		\begin{minipage}[h]{0.4\columnwidth}
			\flushright\includegraphics[scale=1]{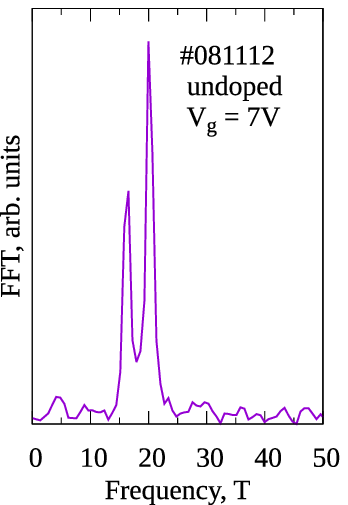} \\
		\end{minipage}
		\begin{minipage}{0.58\columnwidth}
			\flushleft\includegraphics[scale=1]{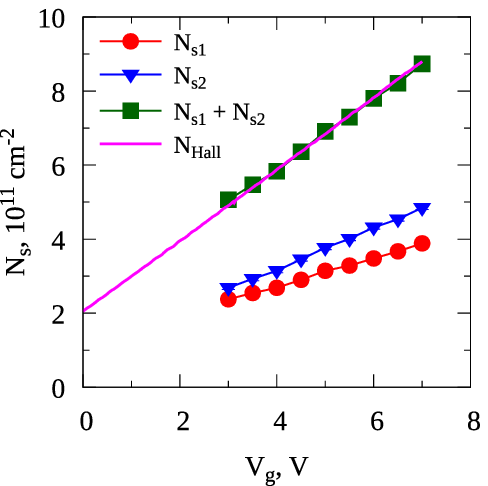} \\
		\end{minipage}
		\vfill
		\begin{minipage}[h]{1\columnwidth}
			\begin{tabular}{p{0.45\linewidth}p{0.6\linewidth}}
				(c) & (d)\\
			\end{tabular}
		\end{minipage}
		\begin{minipage}[h]{0.45\columnwidth}
			\flushright\includegraphics[scale=1]{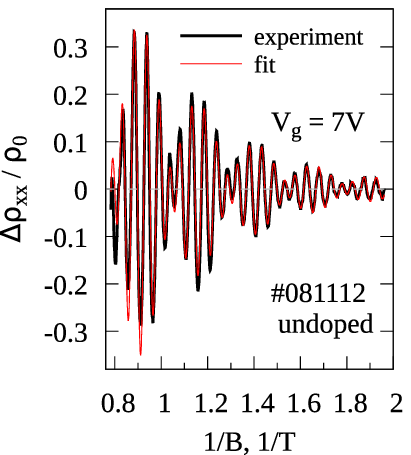} \\
		\end{minipage}
		\begin{minipage}{0.53\columnwidth}
			\flushright\includegraphics[scale=1]{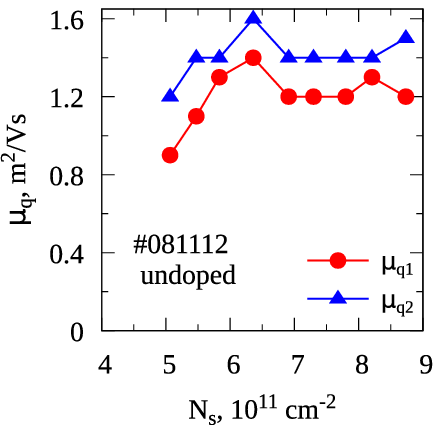} \\
		\end{minipage}
		\caption{Results obtained for undoped 22\,nm HgTe quantum well \#081112: (a)~fast Fourier transformation of $\rho_{xx}(B^{-1})$ at gate voltage $V_g=7$\,V. (b)~Electron concentrations $N_{s1}$ (red circles) and $N_{s2}$ (blue triangular) and their sum (green squares) obtained from Shubnikov--de Haas oscillations and total electron concentration $N_s$ obtained from Hall measurements (pink line) versus gate voltage. (c)~The oscillatory resistivity part $\Delta \rho_{xx}$ normalized to the monotone resistivity part $\rho_0$ versus inverse magnetic field. Black line shows the result obtained experimentally at $V_g = 7$\,V while red line is the fitting curve calculated by Exp.~(\ref{eq:fit}). (d)~Quantum mobilities $\mu_{q1}$ and $\mu_{q2}$ versus total electron concentration.}
		\label{pic:081112_four}
	\end{figure}

	\begin{figure}
		\flushleft\includegraphics[scale=1.2]{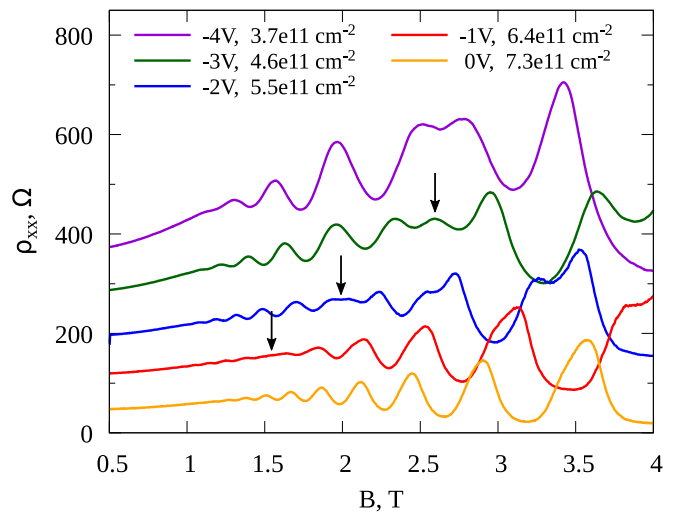} \\
		\caption{Longitudinal resistivity $\rho_{xx}$ dependences on magnetic field $B$ at top gate voltages $V_g$ from 0 to -4\,V obtained for symmetrically \textit{n}-doped 18\,nm HgTe quantum well \#130213. Arrows indicate oscillation beatings.}
		\label{pic:130213_ROxx}
	\end{figure}
	
	\begin{figure}
		\begin{minipage}[h]{1\linewidth}
			\begin{tabular}{p{0.38\linewidth}p{0.26\linewidth}}
				(a) & (b) \\
			\end{tabular}
		\end{minipage}
		\begin{minipage}[h]{0.4\columnwidth}
			\centering\includegraphics[scale=1]{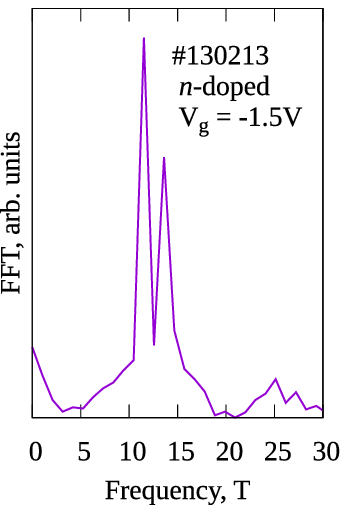} \\
		\end{minipage}
		\begin{minipage}{0.58\columnwidth}
			\flushright\includegraphics[scale=1]{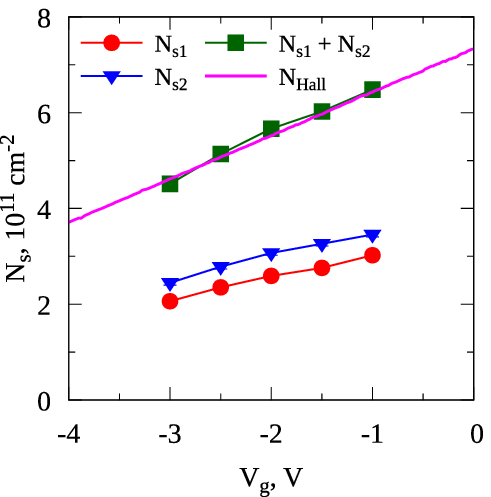} \\
		\end{minipage}
		\vfill
		\begin{minipage}[h]{1\columnwidth}
			\begin{tabular}{p{0.45\linewidth}p{0.33\linewidth}}
				(c) & (d)\\
			\end{tabular}
		\end{minipage}
		\begin{minipage}[h]{0.45\columnwidth}
			\flushleft\includegraphics[scale=1]{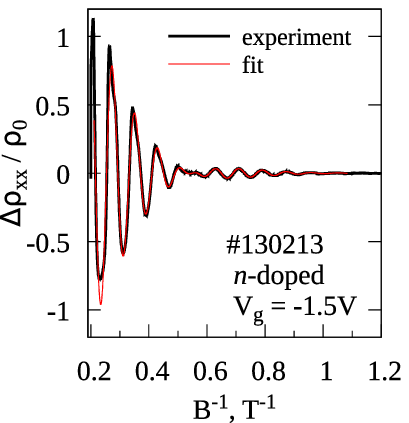} \\
		\end{minipage}
		\begin{minipage}{0.53\linewidth}
			\flushright\includegraphics[scale=1]{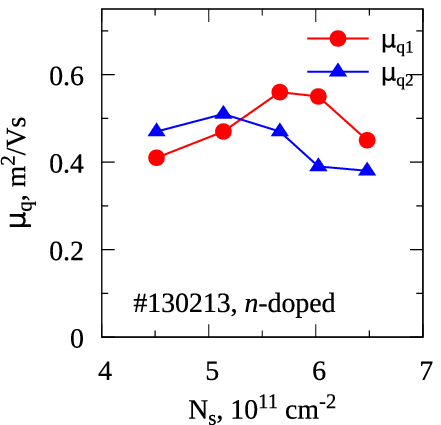} \\
		\end{minipage}
		\caption{Results obtained for symmetrically \textit{n}-doped 18\,nm HgTe quantum well \#130213: (a)~fast Fourier transformation of $\rho_{xx}(B^{-1})$ at gate voltage $V_g=-1.5$\,V. (b)~Electron concentrations $N_{s1}$ (red circles) and $N_{s2}$ (blue triangular) and their sum (green squares) obtained from Shubnikov--de Haas oscillations and total electron concentration $N_s$ obtained from Hall measurements (pink line) versus gate voltage. (c)~The oscillatory resistivity part $\Delta \rho_{xx}$ normalized to the monotone resistivity part $\rho_0$ versus inverse magnetic field. Black line shows the result obtained experimentally at $V_g = -1.5$\,V while red line is the fitting curve calculated by Exp.~(\ref{eq:fit}). (d)~Quantum mobilities $\mu_{q1}$ and $\mu_{q2}$ versus total electron concentration.}
		\label{pic:130213_four}
	\end{figure}
	
	Let us consider our results obtained for the undoped structures first. In Fig.~\ref{pic:081112_ROxx} longitudinal  resistivity $\rho_{xx}$ as a function of magnetic field $B$ is shown for top gate voltages $V_g$ from 0 to 7\,V. Due to good sample quality Shubnikov--de Haas oscillations are already seen at 0.4\,T. The key experimental result is an appearance of oscillation beatings at gate voltage $V_g>3$\,V, whereas at $V_g=0$\,V resistivity oscillations are homogeneous. The oscillation beatings give an evidence of presence of two carrier types in the system with close concentrations. Fourier analysis of resistivity dependence on inverse magnetic field $\rho_{xx}(B^{-1})$ with monotone background removed indeed shows two nearby peaks (see Fig.~\ref{pic:081112_four}\,(a)). From the Fourier analyzes two electron concentrations $N_{s1}$ and $N_{s2}$ can be straight calculated by $N_{si} = ef_i/h$, where we denote by $f_1$ and $f_2$ the lower and upper frequency positions of the Fourier peaks correspondingly. Note the above expression is written for spin non-degenerate electrons, this is justified since at considering gate voltage range only the first conduction subband is occupied.
	
	Although the Fourier analysis enables finding electron concentrations reasonably precisely, we found more accurate getting the frequencies from fitting of Shubnikov--de Haas oscillations by Lifshits - Kosevich formula~\cite{Lifshits1958,Coleridge1989,Fletcher2005}:	
	\begin{align}
	{\Delta \rho_{xx} \over \rho_0} = \sum_{i=1,2}A_i D(X) \text{exp} \left( {-\pi \over \mu_{qi}B} \right) \text{cos} \left( {2\pi f_i \over B} + \phi_i \right),
	\label{eq:fit}
	\end{align}
	where $\rho_0$ is the monotone resistivity part and $\Delta \rho_{xx} = (\rho_{xx} - \rho_0)$ is the oscillatory part; $D(X) = X/\text{sinh}(X)$ is the thermal damping factor with $X=2\pi^2 k_B T/\hbar \omega_c$, $k_B$ being Boltzmann constant and $\omega_c$ being cyclotron frequency; $\mu_{qi}$ are the quantum mobilities; $A_i$ and $\phi_i$ are some constants.
	
	Before fitting the experimental curves we first removed any residual background, which we extracted from the initial curves by Fourier filtering. $A_i$, $\phi_i$, $\mu_i$ and $f_i$ were used as fitting parameters. We used frequencies achieved from Fourier analysis (see Fig.~\ref{pic:081112_four}\,(a)) as starting frequency values. To increase sensitivity to the low-field data we used the weight of 10 for data points at magnetic field less than $\sim0.7$\,T. The fits were always excellent over the full field range, the example of fitting curve for $V_g = 7$\,V is shown in Fig.~\ref{pic:081112_four}\,(c). Concentrations $N_{s1}$ and $N_{s2}$ obtained from the fitting process described above as functions of gate voltage are shown in Fig.~\ref{pic:081112_four}\,(b). The sum of two concentrations $N_{s1}+N_{s2}$ matches very well with the total concentration $N_s$ obtained from Hall measurements.
	
	An additional advantage of oscillation fitting is obtaining quantum mobilities $\mu_{qi}$, which are shown in Fig.~\ref{pic:081112_four}\,(d) as functions of the total electron concentration $N_s$. $\mu_{q1}$ and $\mu_{q2}$ are almost the same and do not change in a full concentration range from 5 to 9$\times10^{11}$\,cm$^{-2}$, also they are more than one order smaller than the transport mobility shown in~Fig.~\ref{pic:sample}\,(b). The difference between transport and quantum mobilities implies presence of long-range scattering, which might be electron density inhomogeneities.
	
	The experimental results for symmetrically \textit{n}-doped quantum well \#130213 are shown in Fig.~\ref{pic:130213_ROxx} and~\ref{pic:130213_four}. Fig.~\ref{pic:130213_ROxx} shows longitudinal  resistivity dependence on magnetic field $\rho_{xx}(B)$ measured at top gate voltages $V_g$ from 0 to -4\,V. Here oscillations are also homogeneous at zero gate voltage while at $V_g<-1$\,V a beating in the oscillations arises providing two peaks in Fourier transformation of $\Delta\rho_{xx}(1/B)$ (see Fig.~\ref{pic:130213_four}\,(a)), $\Delta\rho_{xx}$ is again the oscillatory part of $\rho_{xx}$. Since electron mobility in these structures is smaller than in the undoped ones (see Fig.~\ref{pic:sample}\,(b)), oscillations arise only at $B\sim1$\,T. Together with the elimination at large $B$ by Zeeman splitting it enables only one beating being resolved. Since the beating shifts to larger fields with decreasing gate voltage at $V_g<-3$\,V it disappears due to overlapping with Zeeman splitting.
	
	We performed the same data processing procedure for the sample \#130213 as we did it for \#081112. While fitting the experimental curves by Eq.~(\ref{eq:fit}) we also used the weight of 10 for data points at magnetic field less than $1.5-2$\,T to increase sensitivity to the low-field data. We were succeed to fit all the curves well over the full field range (see as example Fig.~\ref{pic:130213_four}\,(c)), the sum of two concentrations obtained from fitting is in agreement with Hall measurements (see Fig.~\ref{pic:130213_four}\,(b)). Quantum mobilities shown in Fig.~\ref{pic:130213_four}\,(d) are as well as the quantum mobilities in the undoped structure almost the same, do not change in a presented concentration range and one order smaller than the transport mobility.

	\section{Discussion}

	\begin{figure}[t]
		\centering\includegraphics[scale=0.55]{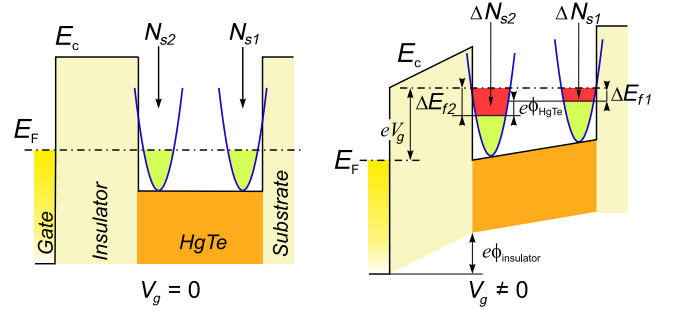}
		\caption{Simplified band diagram and electron distribution over surface states for gate voltages $V_g=0$ (the left panel) and $V_g>0$ (the right panel).}
		\label{pic:BandDiagramm}
	\end{figure}
	
	Beating pattern of Shubnikov--de Haas oscillations at high gate voltages, while at $V_g=0$ the oscillations are homogeneous, in both symmetrically doped and undoped QWs, indicates the origin of the spin splitting being asymmetry of the QW profile, changing with $V_g$. Let us first demonstrate that the difference in the electron concentrations extracted from the ShdH oscillations can be qualitatively explained by a toy electrostatic model involving the surface states at QW interfaces. This model was previously proposed for wide HgTe quantum wells~\cite{Kozlov2016}, and here we briefly repeat its derivation.
	
	As for the relative changes in the concentrations, the initial conditions are not important, therefore, for simplicity, we assume electron concentrations on the top and bottom surfaces being the same at zero $V_g$. Fig.~\ref{pic:BandDiagramm} schematically shows simplified band diagrams and electron distribution over the surface states for a structure with metallic top gate at zero and positive gate voltages. In the absence of gate voltage, the Fermi level remains the same across the structure. When gate voltage is applied, the Fermi level differs in the metallic gate and QW layer by $eV_g$, where $e$ is the elementary charge. Since the left surface is closer to the gate, it partially screens the gate potential from the right surface. The change of electron concentration $\Delta N_{s2}$ at the left surface exceeds thus its changing $\Delta N_{s1}$ at the right one. In their turn, the difference in the concentrations induces an additional electrical potential growth $e\phi_{HgTe}$ between left and right surfaces, while the Fermi level over the QW layer remains constant. The difference in the concentrations can be written as $\Delta N_{si}=\Delta E_{fi}\times D_i$, where $D_i$ ($i=1,2$) is the density of states and $\Delta E_{fi}$ is the local change of the Fermi energy for the right (1) and left (2) surface states. $\Delta E_{f1}$ and $\Delta E_{f2}$ are connected thus as $\Delta E_{f2}=\Delta E_{f1}+e\phi_{\text{HgTe}}$. The potential difference between the two surface states can be evaluated from the charge neutrality and the Gauss's law as $\phi_{\text{HgTe}} = \mathcal{E}_{\text{HgTe}}d_{eff}=e\Delta N_{s1}d_{eff}/\epsilon_{\text{HgTe}}\epsilon_0$, where $d_{eff}$ is the effective distance between the opposite surface states and $\mathcal{E}_{\text{HgTe}}$ is electric field in the well. Here, we neglect a distortion of the QW profile from the linear dependence caused by distribution of charge carriers in the bulk of QW layer. Finally, we find
	\begin{align}
		\Delta N_{s2} / \Delta N_{s1} = D_2 / D_1 + e^2 d_{eff} D_2 / \epsilon_{\text{HgTe}} \epsilon_0.
	\label{eq.dNdN}
	\end{align}
	The effective distance between the surface states $d_{eff}$ can differ from the QW width due to localization of the  surface states wave-functions not exactly on the boundaries of HgTe layer. In addition, the QW width in our samples is comparable with the scale of surface states localization~\cite{Dantscher2015} to exclude the interaction between electrons at different boundaries. Parameter $d_{eff}$ can be evaluated by fitting experimental value of $\Delta N_{s2}/\Delta N_{s1}\simeq(dN_{s2}/dV_g) / (dN_{s1}/dV_g)$ with Eq.~(\ref{eq.dNdN}). It gives $d_{eff}=9$~nm for the sample \#081112 with $(dN_{s2}/dV_g)/(dN_{s1}/dV_g)=1.43$ (see Fig.~\ref{pic:081112_four}), which looks very reasonable for given QW width.

	\begin{figure}[t]
	\begin{tabular}{p{0.51\linewidth}p{0.45\linewidth}}
		(a) & (b) \\
	\end{tabular}
	\begin{minipage}[h]{0.58\linewidth}
		\centering\includegraphics[scale=1]{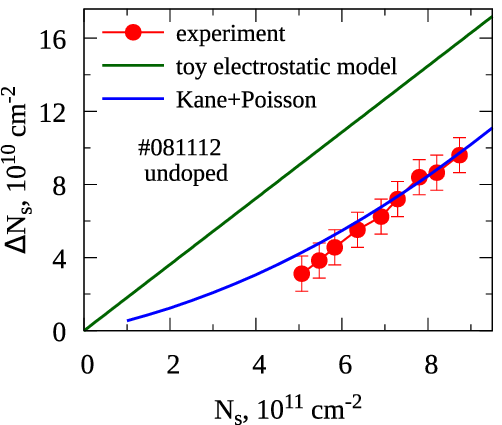}
	\end{minipage}
	\begin{minipage}[h]{0.38\linewidth}
		\centering\includegraphics[scale=1]{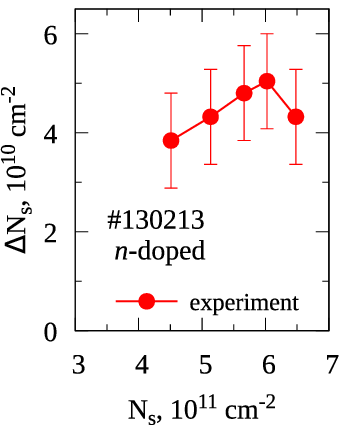}
	\end{minipage}
	\begin{minipage}[h]{1\linewidth}
		\begin{tabular}{p{0.45\linewidth}p{0.5\linewidth}}
			(c) & (d) \\
		\end{tabular}
	\end{minipage}
	\vfill
	\begin{minipage}[h]{0.48\linewidth}
		\centering\includegraphics[scale=1]{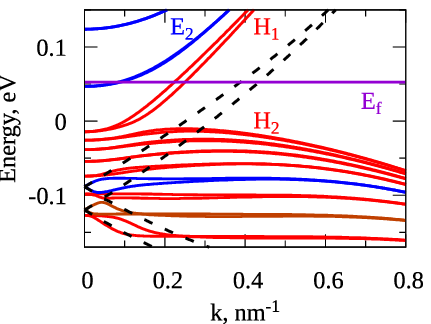}
	\end{minipage}
	\begin{minipage}[h]{0.5\linewidth}
		\flushright\includegraphics[scale=1]{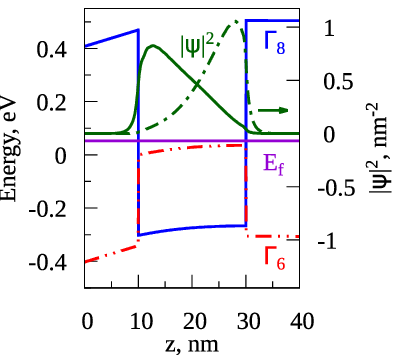}
	\end{minipage}
	\caption{(a) and (b) shows the difference between electron concentrations $\Delta N_s = N_{s2} - N_{s1}$ as a function of total concentration $N_s$ obtained experimentally (red circles) for samples \#081112~(a) and \#130213~(b). In (a) green line corresponds to calculations within toy electrostatic model, while blue line shows self-consistent calculations of Poisson and Schr\"odinger equations with eight-band Kane model Hamiltonian. (c) and (d) shows results of the self-consistent calculations of Poisson and Schr\"odinger equations for electron concentration $N_s = 9\times10^{11}$\,cm$^{-2}$. All electrons are assumed coming to the well due to top gate voltage. (c)~shows the energy spectrum, where black dashed lines correspond to surface states without hybridization with heavy holes. (d)~shows HgTe quantum well potential profile (blue and red lines are $\Gamma_8$ and $\Gamma_6$ bands correspondingly) and squared absolute values of wave functions of electron states at the Fermi level (green lines). }
	\label{pic:dNsWF}
	\end{figure}

	Let us obtain the expression for the difference in electron concentrations at two different surfaces $\Delta N_s = N_{s2}-N_{s1}$ as a function of the total concentration $N_s$. Now, the initial distribution of electrons over the structure becomes important. For simplicity, we assume that $N_s=0$ for symmetric QW profile at $V_g=0$, and all electrons at non-zero $V_g$ come to the HgTe layer due to the top gate voltage. 
	Thus, from $N_{si}=\Delta N_{si}$ and $N_s = N_{s1}+N_{s2}$, we get linear dependence of $\Delta N_s$ on $N_s$:
	\begin{align}
		\Delta N_s = {\Delta N_{s2} / \Delta N_{s1} - 1 \over \Delta N_{s2} / \Delta N_{s1} + 1}N_s.
	\label{eq.dNsNs}
	\end{align}

	Fig.~\ref{pic:dNsWF}\,(a) provides a comparison between experimental data and estimation within our toy electrostatic model (presented by green curve) for the undoped sample \#081112. Here, we used $\epsilon=20$, $d_{eff}=9$\,nm and $D_2=D_1 = m^{*}/2\pi \hbar^2$ valid for parabolic dispersion of the surface states. The latter holds since hybridization with heavy holes modifies the band dispersion of the surface states, making it close to parabolic. From cyclotron resonance measurements~\cite{Kvon2008} the effective mass of the surface states was obtained equal to $m^*\approx0.026 m_0$, with $m_0$ being free electron mass.

	Our toy electrostatic model is seen perfectly reproducing the slope of the experimental behavior of $\Delta N_s(N_s)$. Moreover, it can fit experimental data if one assumes the residual concentration of $4\times10^{11}$\,cm$^{-2}$ in the absence of gate voltage. Note that this value is twice higher than it was measured for the sample \#081112 at $V_g=0$ (see Fig.~\ref{pic:081112_four}). The difference between theoretical estimation and experimental values gives the evidence of the importance of microscopic details of the surface states, which were completely ignored within our toy model.
	
	Therefore, we also perform self-consistent calculations of Poisson and Schr\"odinger equations with 8-band \textbf{k$\cdot$p} Hamiltonian~\cite{Krishtopenko2016}. These calculations take into account all microscopic details of the surface states and thus allow obtaining a realistic QW profile. As it is done for a toy electrostatic model, here we also assume that all electrons at non-zero $V_g$ come to the HgTe layer due to the top gate. At the final iteration of solving self-consistently Poisson and Schr\"odinger equations, we obtain energy dispersions $E(\textbf{k})$ ($\textbf{k}$ is a quasimomentum in the QW plane). Then, for a given value of $N_s$, we find the position of Fermi level and obtain the values of Fermi wave-vectors $k_1$ and $k_2$. Finally, we find electron concentrations by $N_{si} = k_i^2/4\pi$. Theoretical values of $\Delta N_s(N_s)$ found from self-consistent calculations are shown in Fig.~\ref{pic:dNsWF}\,(a) by blue curve and are in a good agreement with the experimental data. 

	Fig.~\ref{pic:dNsWF}\,(c) provides an energy dispersion of the surface states at $N_s = 9\times 10^{11}$\,cm$^{-2}$, where they are represented by $H_1$ subband due to hybridization with the states of heavy-hole band. Surface state connection with the $H_1$ subband is also supported by Fig.~\ref{pic:dNsWF}\,(d). The figure shows theoretical QW profile and wave-functions of the states at the Fermi level (see green curves). Spin-split states corresponding to $k_1$ and $k_2$ wave-vectors are clearly seen to localize at the opposite boundaries of HgTe QW. Large overlapping between the surface states in our samples also explains only qualitative agreement of the experimental data with our toy electrostatic model. We note that hybridization of the surface states with the heavy-hole band is partially included in the toy model by using expression for the density of states $D= m^{*}/2\pi \hbar^2$, which is inherent for parabolic spectrum. The dashed black curves show dispersion of the surface states neglecting hybridization with the heavy holes. The surface states mixing with the $|\Gamma_8,\pm3/2\rangle$ band is indeed seen transforming the linear dispersion of surface states into parabolic. Interestingly, the spin splitting of the surface states is significantly suppressed if the hybridization is included.

	$\Delta N_s (N_s)$ obtained experimentally for the \textit{n}-doped structure \#130213 is shown in Fig.~\ref{pic:dNsWF}\,(b). This pattern contradicts our expectations of the spin splitting increasing with the absolute gate voltage value and decreasing thus electron concentration. The reason is likely the presence of only one beating in the ShdH oscillations and thus less precise electron concentration determination. As seen in Fig.~\ref{pic:130213_four}\,(b) it is not crucial for determination of the total electron concentration however seems significant for that of electron concentration difference.
	
	\section{Conclusion}
	
	To sum up we have investigated Rashba-like spin splitting of the conduction $H1$ band in 18--22\,nm HgTe quantum wells. Beating pattern of Shubnikov--de Haas oscillations, arising with applying top gate voltage in both undoped and symmetrically \textit{n}-doped structures, provides two close electron concentrations. We have qualitatively described the evolution of the difference between these concentrations with gate voltage by a toy electrostatic model involving electron states localization at the well interfaces. The quantitative agreement between the experimental data and theoretical calculations was achieved by self-consistent solving Poisson and Schr\"odinger equations with eight-band \textbf{k$\cdot$p} Hamiltonian, which takes into account microscopic details of the surface states omitted in our toy model. Comparison of the toy electrostatic model with the rigorous self-consistent calculations clearly shows large spin-splitting of $H1$ subband in the HgTe quantum wells being due to the surface nature of its states.	
	
	\section{Acknowledgments}
	
	This work was supported by the Russian Foundation for Basic Research (projects no. 17-52-14007) and Government of Novosibirsk region (projects no. 17-42-543336).

\end{document}